\documentclass[%
aip,
jmp,%
twocolumn,
amsmath,
amssymb,%
reprint,
author-year,
author-numerical,%
]{revtex4}
 
\usepackage{graphicx}
\usepackage{dcolumn}
\usepackage{bm}
\usepackage{color}
 
\usepackage[
   colorlinks,
   linkcolor=blue,
   urlcolor=blue,
   citecolor=blue,
   a4paper=true,
   ]{hyperref}

\begin{document}
 
\preprint{AIP/123-QED}
 
\title{Structure and interactions of ultracold Yb ions and Rb atoms}

\author{H. D. L. Lamb}
\email{hlamb01@qub.ac.uk.}
\author{J. F. McCann}%
\email{j.f.mccann@qub.ac.uk}
\author{B. M. McLaughlin}%
\email{b.mclaughlin@qub.ac.uk}
\affiliation{
Centre for Theoretical Atomic, Molecular and Optical Physics,
School of Mathematics and Physics, Queen's University Belfast,
Belfast BT7 1NN, Northern Ireland, UK.
}%
 
\author{J. Goold}
\affiliation{Clarendon Laboratory, University of Oxford, UK,  \\  
and Department of Physics, University College Cork, Ireland.
}%
 
\author{N. Wells}
\author{I. Lane}%
\affiliation{
Innovative Molecular Materials  Group,
School of Chemistry and Chemical Engineering, Queen's University Belfast,
Belfast BT9 5AG, Northern Ireland, UK.
}%
 
 
 
\date{\today}
 
\begin{abstract}
In order to study ultracold charge-transfer processes in hybrid atom-ion traps,  we have mapped out the 
potential energy curves and molecular parameters for several low lying states of the Rb, Yb$^+$ system. 
We employ both a multi-reference configuration interaction (MRCI) and a full configuration interaction (FCI) 
approach.  Turning points, crossing points, potential minima and spectroscopic molecular constants 
are obtained for the  lowest five molecular states.  Long-range parameters, including  
the dispersion coefficients are estimated from our {\it ab initio} data.
The  separated-atom ionization potentials and atomic polarizability  of the ytterbium atom ($\alpha_d=128.4$ atomic units)
are in good agreement with experiment and previous calculations. We present some dynamical calculations for 
 (adiabatic) scattering lengths for the two lowest (Yb,Rb$^+$) channels that were carried out in our work. 
 However, we find that the pseudo potential approximation 
is rather limited in validity, and only applies to nK temperatures. 
The adiabatic scattering lengths for both the triplet and singlet channels 
indicate that both are large and negative in the FCI approximation. 
\end{abstract}

 
\maketitle 
 
\section{\label{sec:level1}Introduction}
As a gas is cooled down towards micro-Kelvin temperatures, the quantum nature of the interactions
begins to dominate.  With increasing de Broglie wavelength,  the long-range tail of the  potential
plays a vital role in determining the nature of the interactions. In particular whether the elastic pair-wise interaction
is, in the limit of zero temperature, attractive or repulsive ~\cite{foot04,cote98}. 
The lower the incident energy of the ion-atom pair,
the larger the distances for which these potentials influence the phase shift.  This is critical in terms
of the ultracold regime as to whether cooling, trapping and degeneracy can occur.  It is extremely
important in view of  potential  applications in exploring the fundamental process of ultracold charge transfer ~\cite{fior98,stwa99,wein99,kohl12}.
 
Interest has developed in expanding the range of quantum systems that can be trapped and manipulated on the quantum
scale. Hybrid ion-atom systems are of great interest ~\cite{grie09} since these are inherently strongly-interacting systems 
with a longer-range potential, and inelastic processes can be studied. Recently these systems have been 
explored considering two-body collisions, in which both collision partners are translationally 
cold~\cite{zipk10b}, and on the many-body level~\cite{zipk10a}, where the sympathetic cooling 
of the ion with ultracold atoms was observed. The study of these systems in the quantum regime can be applied to hybrid atom-ion
devices~\cite{idzi07} and, in addressing fundamental many-body effects of ionic impurities such as
mesoscopic molecule formation~\cite{lukin} and density fluctuations~\cite{gool10}. These devices also 
afford a unique opportunity to study reactive collisions (ultracold chemistry)  under controlled conditions, for 
example when external electric fields can be applied to modify the reaction rates/cross sections ~\cite{zipk10a}. 
Unlike binary cold collisions between ground state neutral atoms, which are only elastic or inelastic in nature, 
reactive collisions (charge transfer) are a feature of Yb ions immersed in a gas of trapped alkali atoms 
as the products Yb + Rb$^{+}$ are the thermodynamically favored species. Consequently, curve crossings 
between electronic states, again absent in the neutral systems studied to date, 
can play a significant role in determining the collision dynamics. 
 
Ultracold neutral atom interactions are characterized by pure $s$-wave scattering mediated at long-range by the
dispersion forces  \cite{hirs54}. Conversely, a bare ion creates a polarization force directly and hence
the effective cross section is larger with significant contributions from  higher-order partial waves  ~\cite{cote00}. 
Indeed the usual effective range expansion is modified by logarithmic terms in the wavenumber expansion~\cite{omal61}.
In the last few years theoretical studies of ultracold atom-ion collisions included the investigation 
of the occurrence of magnetic Feshbach resonances with a view to examining the tunability of the 
atom-ion interaction focusing on the specific $^{40}{\rm Ca}^{+} - {\rm Na}$ system~\cite{idzi09}, 
and calculations of the single-channel scattering properties of the Ba$^+$ ion with the Rb 
neutral atom~\cite{kryc10} which suggest the possibility of sympathetic cooling of the barium 
ion by the buffer gas of ultracold rubidium atoms with a considerable efficiency.
 
In recent experiments ~\cite{zipk10a,zipk10b,denschlag}, a single trapped ion of $^{174}$Yb$^+$ in a Paul trap was immersed in a condensate of
neutral $^{87}$Rb atoms confined in a magneto-optical trap.  A study of charge transfer cross sections showed that the 
simple classical Langevin model was inadequate to explain the reaction rates ~\cite{zipk10a}. However, very little is known about the microscopic
ultracold binary interactions between this ion and the rubidium atom. In particular, the potential energy curves and couplings 
are not known with any  accuracy. Thus the experimental study of the quantum coherence of charge transfer ~\cite{zipk10a} was based on schematics 
of the energy curves.  This prompted our in depth investigations. 
Our initial aim was to map the potential energy curves to establish the adiabatic states and  the static properties of the molecular
ion, in particular the turning points, potential minima, and crossing points of the lowest molecular energies. In addition to this, 
the dissociation energies and molecular constants would provide useful spectroscopic data for further investigation. We have 
made a preliminary estimation of the pseudopotential which approximates the ultracold interaction.
This information is of great importance for modeling ultracold charge transfer, and in particular the quantum character of chemical reactivity 
and thus develop insights into ultracold quantum controlled chemistry, for example when external fields 
are applied to influence the reaction rates and reaction channels ~\cite{zipk10a}. Of course, the presence of a bare charge in a dilute gas exposes 
many-body physics features such as exciton and polariton dynamics, which are also of great interest.  
It is also of great interest for laser manipulation of the collision to prevent losses 
through charge transfer or create translationally-cold  trapped molecular ions via photoassociation.
 
\section{Electronic structure calculation}
 
Beyond the calculation of ionization potentials and dissociation energies, 
one of the more challenging tasks for a quantum chemistry package is the calculation of  polarizabilities and
dispersion forces. Since these terms dominate  at asymptotic long-range and low energies, 
their accuracy is paramount in  obtaining a solid foundation for  scattering calculations \cite{mitr10}.
In atomic units, the asymptotic singly-charged ion-atom potential has the form~\cite{hirs54,cote00}
\begin{equation}
  V(R)=V_{\infty}- \frac{1}{2}\left[ \frac{\alpha_d}{R^4} +\frac{ C_6}{R^6}+ \frac{C_{8}}{R^8} \right],
\label{mult}
\end{equation}
where  $\alpha_{d}$ is the  dipole polarizability of the neutral atom and
where $C_6$ and $C_8$ are respectively the quadrupole and octupole polarizabilities, 
and $V_{\infty}$ is the asymptotic limit (ionization potential). $R$ is the ion-atom internuclear separation.
Of course, the ionization potential and the other coefficients will depend on the electronic 
configuration and thus on the electronic state in question. The primary interest, for 
cold atom physics, is the ground-state. However, for temperatures in the mK regime, 
the excited states have an important role in energy exchanges through collisions.   
A recent review of  existing theoretical methods for ground-state polarizability discusses 
their relevance to  cold-atom physics~\cite{mitr10}.
Calculations of polarizability  for molecules with heavy atoms, even restricting interest to 
the ground state,  
become increasingly difficult, partly because of  relativistic corrections. However, as is the case
of Rb,  electron correlation is in practice more problematic. For example,  a basic
non-relativistic Hartree-Fock calculation \cite{lim99} gives $\alpha_{d} \approx 522$ (atomic units). It  requires 
the power of the coupled-cluster expansions to account fully for correlation. The
CCSD(T) method gives \cite{lim99} $\alpha_{d} \approx 352$(a.u.) (non-relativistic), while 
relativistic corrections take the value to, $\alpha_{d} \approx 324$(a.u.). At this point, the 
theory is  consistent with experiment:  $319 \pm 6$ \cite{molo74}. However, excited state 
calculations are more problematic. 

In our approach we employed the MOLPRO~\cite{wern10} suite of {\it ab initio} quantum chemistry codes 
(release MOLPRO 2010.1) to perform all the calculations for this diatomic system (Rb,Yb)$^+$ to 
obtain the potential energy curves (PEC's) as a function of bond length. 
We used a combination of multi-reference configuration interaction (MRCI) and full configuration interaction (FCI) approximations. 
The {\it ab initio} potential energy curves for this diatomic system are calculated using effective core potentials (ECP) as a basis set for each atom, which allows for 
scalar-relativistic effects to be included explicitly. The scalar-relativistic effects are included by adding the corresponding terms 
of the Douglas-Kroll Hamiltonian to the one-electron integrals.
For the short-range interactions for the molecular electronic states we used the 
non-relativistic complete-active-space self consistent field (CASSCF)/MRCI method~\cite{wern85,know85,wern88,know88} 
within the MOLPRO~\cite{wern10} {\it ab initio} quantum chemistry suite of codes.  

We used the Stuttgart basis sets and effective core potentials (ECP's)
with core-polarization potentials (CPP) for Rb and Yb, respectively the ECP68MDF potential  for Yb,
and the ECP36SDF potential for Rb. The inner shell electrons are modeled using these effective core potentials 
(ECP68MDF) for ytterbium~\cite{wang98}, and (ECP36SDF) for rubidium~\cite{szen82,fuen83}. 
We started at a bond separation of $R = 30$ $a_0$ and moved in to $R=5$ $a_0$. 
At $R=30$ $a_0$, we perform a self-consistent field Hartree-Fock SCF-HF calculation on
the closed-shell YbRb molecular ion system to obtain
a starting wave function. This is the starting configuration for the multireference
configuration-interaction MRCI calculation
performed in the appropriate molecular symmetry group. We used
an active space of \{6a$_1$, 3b$_1$, 3b$_2$, 0a$_2$\} in the C$_{2v}$ Abelian point group with no closed orbitals.
 We then perform a multi-configuration self-consistent field calculation MCSCF.
The multi-configuration self-consistent field (MCSCF) calculation was performed on the lowest five electronic states.
These results were used as the initial wavefunction set for a multi-reference configuration interaction (MRCI) 
calculation to capture the dynamic electron correlation. 

Only the valence shells were included in the determination of the electronic correlation energy.
As a test of the basis set employed, we conducted equivalent calculations on the neutral YbRb molecule 
yielding values as expected consistent with those of Meyer and Bohn~\cite{meye09}.
 
\begin{figure}
\includegraphics[width=0.525\textwidth]{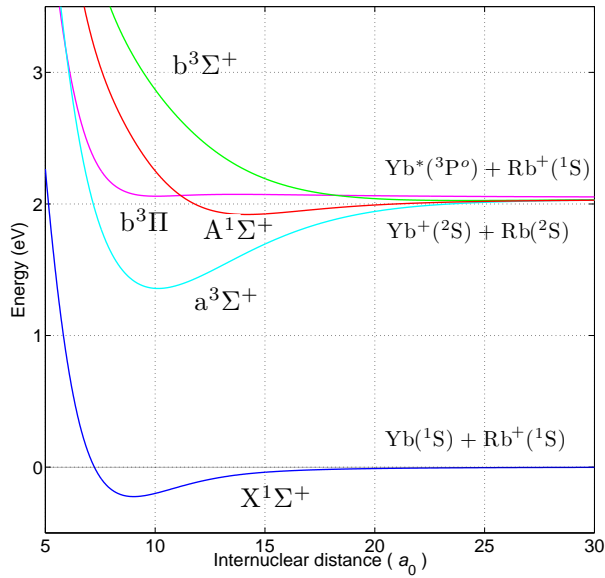}
\caption{Relative electronic energies for the molecular ion ${\rm YbRb}^+$ as a 
               function of internuclear distance, $R$, (MRCI approximation). The ${\rm X}^{1}\Sigma^{+}$ 
                ground state is the Rb$^{+}$ channel, while the lowest energy ionic ytterbium states, 
                the triplet $a ^{3} \Sigma^{+}$ and singlet $A ^{1}\Sigma^{+}$ pair, are nearly degenerate 
                with the excited charge-transfer channels: $ {\rm Rb}^{+}+{\rm Yb}^{*}$.}
\label{fig1}
\end{figure}
 
\begin{table*}
\caption{Comparison between the experimental and calculated molecular asymptotes. 
                The experimental asymptotes are derived from ionization potentials~\cite{camu80,ayma80,megg78,joha61}. 
                The calculated asymptotes are obtained by the fit (2) using the LEVEL code~\cite{lero07}}
\begin{ruledtabular}
\begin{tabular}{lcccc}
Molecular & Estimated $E_\infty^{(i)}$ & Experimental &             $\Delta$E Difference   &  $\Delta$E Difference \\
Symmetry  & (eV) & Asymptotic Limit (eV) &  (eV) & (\%) \\
\hline
X$^1\Sigma^+$ & 0  & 0 & 0 & - \\
a$^3\Sigma^+$ & 2.036 & 2.076 & 0.040 & 1.93 \\
A$^1\Sigma^+$ & 2.036 & 2.076 & 0.040 & 1.93 \\
b$^3\Pi$ & 2.045 & 2.143 & 0.089 & 4.15\\
b$^3\Sigma^+$ & 2.050 & 2.143 & 0.093 & 4.34 \\
\end{tabular}
\end{ruledtabular}
\label{tab1}
\end{table*}
 
\begin{table*}
\caption{Molecular constants for the four lowest states supporting rovibrational bound 
                states in the MRCI approximations.  The equilibrium bond length $r_e$ is in atomic units, the 
               dissociation energy $D_e$ is in eV, both were obtained from the {\it ab initio} data. 
               The rovibrational constants defined in (2) are in cm$^{-1}$.
              Note that the b$^3\Sigma^+$ state is repulsive.}
\begin{ruledtabular}
\begin{tabular}{lccrccc}
Molecular &  $r_e$ & $\omega_e$ &   $\omega_e x_e$ &  $B$ & $D(10^{-9})$& $D_e$\\
Symmetry  &  $a_0$ & cm$^{-1}$   &  cm$^{-1}$  &   cm$^{-1}$ &   cm$^{-1}$ & eV \\
\hline
X$^1\Sigma^+$ & 9.031   & 33.77 & $-0.22$ & 0.01273 & 7.0047 & 0.2202 \\
a$^3\Sigma^+$ & 10.142 & 34.94 & $0.04$ & 0.01010 & 3.3843 & 0.6653 \\
A$^1\Sigma^+$ & 14.362 & 16.807 & $0.38$ & 0.00505 & 2.1440 & 0.1085 \\
b$^3\Pi$ & 10.108 & 15.24 & $-0.21$ & 0.00102 & 16.812 & 0.0061\\
\end{tabular}
\end{ruledtabular}
\label{tab2}
\end{table*}

Since the MRCI calculations do not explicitly include relativistic effects, although this is not
important for the entrance collision channel or the lower Yb ($^1$S) + Rb$^+$ ($^1$S) asymptote as all the
molecular states formed are of $\Sigma^+$ symmetry. This is borne out by the calculated energy of the
asymptotic energies of the a$^3\Sigma^+$ and A$^1\Sigma^+$ states (Table \ref{tab1}).
The asymptotes for the higher $^3\Pi$ and $^3\Sigma^+$ states 
correlate  to the Yb (6s6p~$^3$P$^o$) + Rb$^+$ (4p$^6$~$^1$S) atomic products. The multiplet 
and its associated fine-structure splitting in the triplet (Yb: $^3$P$^o_{0,1,2}$)  is considerable: $\sim$0.3 eV. 
Only a fully relativistic treatment can 
accurately account for the spin-orbit interaction. In a magnetic trap of course the Zeeman splitting and hyperfine structure complicates 
matters further. Nonetheless, in our first analysis of this novel system, we can confidently say 
that a curve crossing will take place between the A$^1\Sigma^+$ and b$^3\Pi$ states
though at an energy above the Yb$^+$ ($^2$S) + Rb ($^2$S) asymptote. Such a crossing will facilitate a charge exchange
reaction as observed in experiment at mK temperatures~\cite{ zipk10b,zipk10a}. We have estimated the 
molecular constants for the four states that support bound rovibrational states.  These constants
are defined by the usual expression, for electronic state (channel) $i$ with asymptotic energy $E^{(i)}_{\infty}$, 
\begin{eqnarray}
E(i,v,J) = E^{(i)}_{\infty} - D^{(i)}_e && \\ \nonumber
	       +\hbar \omega^{(i)}_e \left(v+{\textstyle  \frac{1}{2}}\right) &- \hbar \omega^{(i)}_e x_e \left(v+ {\textstyle \frac{1}{2}} \right)^2 & \\ \nonumber
                + B^{(i)} J(J+1) & - D^{(i)} J^2 (J+1)^2 &  \quad . 
                \label{rovib}
\end{eqnarray}
All the constants were derived from the electronic potentials and were calculated using 
 the LEVEL~\cite{lero07} program (version 8.0). The values  are presented in Table \ref{tab2}. 

\subsection{ Scattering length calculation}
 
In the adiabatic approximation the dynamics occur on decoupled, centrally-symmetric potential energy curves.  
Using conventional notation,  we let $E$ denote the collision energy in the  center-of-mass frame, and 
$m_i$ and $m_a$ the ion and atom masses, respectively. Then, the reduced mass is: 
 $\mu=m_i m_a / (m_i + m_a)$. Using atomic units, unless otherwise stated, 
 the molecular channel $i$, $k^2 = 2\mu ( E - E^{(i)}_{\infty}) \ge $ 0  is the  relative wavenumber squared.  
 The mean-square wavenumber  has the equivalent  temperature, $T=\hbar^2 \left < k^2 \right> / (3k_B \mu)$, 
 where $k_{B}$ is the Boltzmann constant. 
 For this $^{174}$Yb$^+$,$^{87}$ Rb system, we have that $k {\rm (a.u.)} \approx \sqrt{T {\rm (K)}}$. From fig 2 one sees 
 that the upper limit of the energy range, $k= 10^{-4}$ a.u.,  corresponds to $T \approx 10 $ nK.
 
 The radial Schr\"{o}dinger equation  for  the $s$-wave  in the 
 channel/potential $i$ is then:
\begin{equation}
\left[
\frac{d^2}{dR^2} + k^2 - 2\mu V_{i}(R)  \,
\right] \chi_{i}(k,R) ~=~0
\label{scatt}
\end{equation}
where $V_{i} (R)  \sim -\alpha_d/(2R^4)$ as $R \rightarrow \infty$.  
Then the phase shift $\delta_i(k)$ has its usual definition,
\begin{equation}
\chi_{i}(k,R) ~\sim~
{\rm sin} \left(   kR  + \delta_i (k)  \right) ~ , ~  R\rightarrow \infty \quad .
\label{phas}
\end{equation}
At ultracold temperatures, $k \rightarrow 0$,  the effective range expansion for the $s$-wave~\cite{omal61},  takes the form\\
\begin{eqnarray}
k \cot \delta_i (k)  =&   -\dfrac{1}{a_s} +  \dfrac{\pi \mu \alpha_d}{ 3a_s^2}k \qquad  \qquad \qquad   \\ \nonumber
& + \dfrac{4 \mu \alpha_d}{3a_s} k^2 \ln \left( \dfrac{k}{4}\sqrt{\mu \alpha_d} \right) +\mathcal{O}(k^2)
\label{omal}
\end{eqnarray}
We note the logarithmic terms in $k$  as opposed to the usual quadratic term for a short-range force. 
Clearly at zero temperature, the $s$-wave scattering length, $a_s$,  has the usual expression,
\begin{equation}
a_{i,s} = \lim_{k \rightarrow 0} - {\tan \delta_i (k) \over k}.
\label{scatlen1}
\end{equation}
The significance of the scattering length for ultracold gases  is  as a  strength parameter
for a contact two-body pseudopotential that can, in turn, be used in the many-body Hamiltonian appropriate 
for an ion imbedded in a cloud of cold atoms. Thus, for channel $i$, and ion-atom 
separation $ \vec{R}$,  we have the zero-temperature pseudopotential (operating to the right):
\begin{equation}
U_{i}(\vec{R}) = {2 \pi \hbar^2 \over \mu} a_{i, s}\  \delta( \vec{R} ) {\partial \over \partial R} \cdot R \ .
\label{pseudo}
\end{equation}

\noindent 
In general, the pseudopotential can be made energy dependent: $a_{i,s}$ is replaced by $a(k)$, 
where $k$ is the wavenumber.  At distances near the equilibrium bond length, $r_e$, the interaction 
between the ion charge and the dipole moment in the atom results 
in a deep potential well for which the thermal de Broglie wavelength, $\lambda$, is small, with the potential
slowly-varying a semi-classical approximation is justified. At long range, where
the dipole potential is the dominant term, the radial Schr\"odinger equation \eqref{scatt} can 
be solved in closed analytic form. Connecting the
exact asymptotic  solution with the semi-classical approximation at a {\em matching distance} ($R_{c}$)
gives the phase shift in terms of a simple quadrature  \cite{grib93,grib99}.
This, in turn, yields an elegant and simple expression for the scattering length,
\begin{equation}
a_s=-\sqrt{\mu \alpha_d}\ {\rm tan} \left( \Phi - \frac{\pi}{4} \right)
\end{equation}
where
\begin{equation}
\Phi =  \int_{R_0}^\infty{ \sqrt{-2\mu V(R)}}\  dR
\label{phase}
\end{equation}
with $R_{0}$ the zero-energy classical turning point i.e. the smallest value for the solution of,  $V(R_{0})=0$.
The phase \eqref{phase} is the accumulation of the inner (semi-classical) phase as far as the matching radius, $R_{c}$,
and  the outer (asymptotic) phase beyond the matching radius
in the region where the potential is approximately a pure dipole. This can be simply written as  \cite{grib93,grib99}:
$\Phi=\Phi_< + \Phi_>$, where
\begin{equation}
\Phi_< =  \int_{R_0}^{R_c} \sqrt{-2 \mu V(R)} dR, ~~ \Phi_> = \sqrt{\mu \alpha_d}  \frac{1}{R_c}.
\end{equation}
In accordance with Levinson's theorem,  the phase $\Phi$  passes through many cycles of $\pi$ at the
threshold energy. For a dipole potential and within the semi-classical approximation,
the number of bound states is given by
\begin{equation}
n_s = {\rm int} \left[ \Phi/\pi - 3/4 \right] +1.
\end{equation}
 
 The large, but uncertain, value of the phase leads one to the conjecture~\cite{grib93} that
there is equal likelihood that the scattering length is positive or negative, and indeed it may  be infinitely
large in magnitude. This sensitivity of scattering length to phase shift, which in 
turn depends on the interatomic potential amplified by the large reduced mass, 
means that obtaining accurate and reliable theoretical estimates of the
scattering length is extremely difficult. Nonetheless, one of the aims of this paper
is to  make a first estimate of these values.
 
\section{Results}
  
\subsection{Ground state}
 
First, we consider the $X^1\Sigma^+$ electronic potential corresponding to the separated ion-atom asymptotic channel,
Rb$^+$ ($\rm 4p^6~^1S$) + Yb($\rm 6s^2~^1S$). In this case, the ytterbium atom experiences the
long range single charge of the Rb ion, and the electronic potential has the leading order
term of order $R^{-4}$. As a test of the accuracy of the electronic energy curve
we obtained,  we estimated the ytterbium polarizability by curve fitting the potential to $-\frac{1}{2}\alpha_d R^{-4}$ at values
of $R > 30$(a.u.).  Using a least-squares fit over 15 points using a constant and $R^{-4}$ 
as independent variables for which we computed the coefficients corresponding respectively 
to the ionization energy and the polarizability of the atomic species. For the ground electronic 
state $X^1\Sigma^+$, which corresponds asymptotically to Rb$^+$ ($\rm 4p^6~^1$S) and 
neutral Yb(6s$^2$~$^1$S), the hyperpolarizability terms have a small contribution, so
at internuclear distances beyond  30 a$_0$ they may be neglected.
From this approach we find estimates of the polarizability of ytterbium $\alpha_d \approx 128.5$ in both MRCI and FCI approximations, 
which are in suitable agreement with results from previous investigations, see Table \ref{tab3}.
 
In order to obtain estimates for the scattering length, we performed numerical integration of the radial Schr\"odinger equation (3) 
to determine the phase shift using the Runge-Kutta method along with a numerical quadrature (Simpson's rule)  for the semi-classical
approximation.  The numerical integration was initiated just to the left of the classical turning
point and integrated outwards to fit to the form \eqref{phas}.
 
\begin{figure}
\includegraphics[width=.5\textwidth]{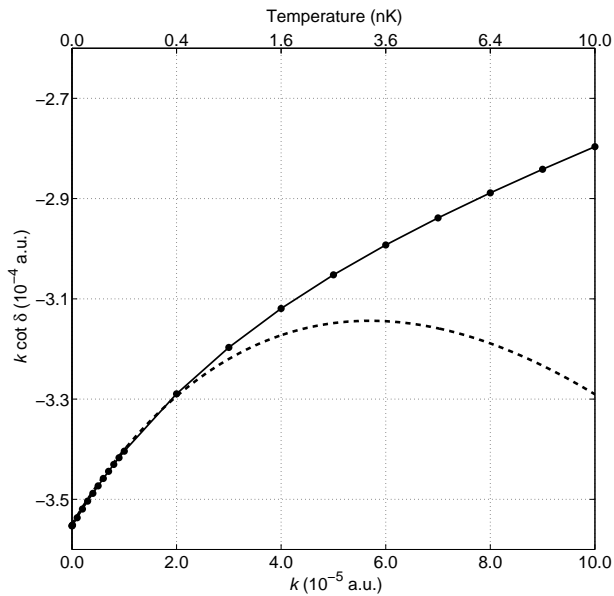}
\caption{ The effective-range function, 
 $k~{\rm cot}~\delta(k)$, in the limit of low energy as a function of the relative wavenumber, $k$, and equivalent temperature, 
for the  X$^{1}\Sigma^{+}$ state in the MRCI approximation. 
                 The solid line is the value obtain by numerical integration. 
                  The dashed curve displays the low-temperate expansion (5). 
                  For this system, it is clear that replacing the scattering by a temperature-independent 
                  pseudo-potential (\ref{pseudo}) is only valid in the nK r\'egime.}
\label{fig2}
\end{figure}
 
The phase shift was determined by numerically integrating the radial Scr\"odinger equation (3) using a simple 4th/5th order Runge-Kutta  scheme 
starting within the classically forbidden region out to the matching radius $R_c \sim  37 a_0$.
This  distance was sufficiently large enough to ensure that beyond this value the potential is then suitably
represented by the multipole expansion (1).  The leading-order dipole term with
the value $\alpha_d=128.4894$, gave $n_ s$ =155 bound states for the MRCI calculation. 
The full configuration interaction calculation gives $n_ s$ =155.  From this we find a large negative scattering length
indicating the presence of a virtual bound state at $E=(2\mu a_s^2)^{-1}$.
Our calculated polarizability (see Table~\ref{tab3} ) falls within the range of previous theoretical and
experimental results. 
 
\begin{table}
\caption{Static electric-dipole polarizability $\alpha_d$ (a.u.) for the
                ground state of Yb (6s$^2$~$^1$S$_0$) compared
                 with previous experimental and theoretical work.}
\begin{ruledtabular}
\begin{tabular}{ll}
$\alpha_d$ & Comment\\
\hline
128.5 & This work (MRCI)\\
128.4 & This work (FCI)\\
$141 \pm 6$ & Dzuba and Derivianko; CI+MBPT\cite{dzub10}\\
$136.4 \pm 4.0$ & Experimental data~\cite{zhan07}.\\
157.3 & Chu {\it et al}; DFT~\cite{chu07} \\
$141 \pm 35$ & Linear response method~\cite{zang80}.\\
\end{tabular}
\end{ruledtabular}
\label{tab3}
\end{table}
 
The de Broglie wavelength increases  as $k\rightarrow 0$. For $k<10^{-3}$ ($T\lesssim 1$ $\mu$K) the collisional properties are 
determined by s-wave scattering. We are already well into the nK regime when $k=10^{-8}$, 
for which convergence of the phase shift to its value in the low energy limit, to reasonable accuracy,  
can be assumed. Integration is carried out to a distance for which the long ranged dispersion 
potentials have a negligible effect on the phase, $k^2 \gg \mu\alpha_d / R^4$. The method of 
extrapolation of the zero-energy wave function is less computationally expensive and can be 
used as a test against the results for very low energy scattering. 
 
The numerically obtained low-energy s-wave phase shifts are in good agreement with those 
from the  low-energy phase shift expansion (5) using the semi-classical scattering 
length ($a_{SC}=2816.7$ in the MRCI approximations) as illustrated in Fig.~\ref{fig2} 
for the range of $ka_s \lesssim 0.3$.
 
The variation in the effective scattering length:
\begin{equation}
 a(k) = - \frac{{\rm  tan}\delta (k)}{k},
 \label{ak}
\end{equation} 
\noindent with relative wavenumber, $k$, is shown in Fig.~\ref{fig3} below. 
We also consider comparison with the linear expansion of Eq. (5), via the binomial approximation, 
which gives in this case,
\begin{equation}
a(k) = a_s  \left( 1- \frac{\pi\mu \alpha_{d}}{3 a_s}k  \right)^{-1} \approx a_s \left( 1+\frac{\pi\mu \alpha_{d}}{3 a_s}k \right).
\label{linear}
\end{equation}

In Fig.~\ref{fig3} the numerical calculation (solid line with circles) deviates both from the linear and nonlinear expansions, 
even at these extremely low temperatures. Based on these preliminary calculations, 
current experiments, operating in the $\mu$K range, would have 
great difficulty in estimating the scattering length. The linear approximation is adequate for temperatures  $T < 1~$nK, while the 
expansion (5) breaks down at   $T= 1~$nK, as 
higher-order terms then become significant.
 
\begin{figure}[t]
\includegraphics[width=.5\textwidth]{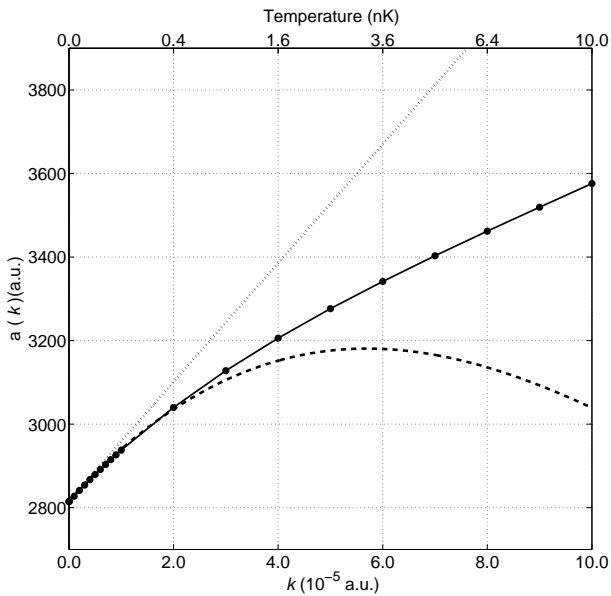}
\caption{The energy-dependent scattering length, $a(k)$, defined by \eqref{ak}, for
                the ground state potential curve ($X^{1}\Sigma^{+}$) in the MRCI approximations.
                The solid curve is the result of numerical integration,     the   dotted  line is the linear approximation \eqref{linear} 
                and the dashed curve in the nonlinear expansion (5). Even over this ultracold (nK) temperature range  the 
                variation of the effective scattering length is significant. }
\label{fig3}
\end{figure}

\subsection{Excited states}
 
In ion trap experiments, the primary interaction will be
the Yb$^+$(6s~$^2$S) ion colliding with the  neutral Rb (5s~$^2$S) which may occur
in the   A$^1\Sigma^+$ singlet or the a$^3\Sigma^+$ triplet state.  We see (Fig~\ref{fig1}) that
the triplet has a lower minimum and shorter bond length, and because of the
statistical weight is the more important channel. In the separated-atom limit, this channel
  lies 2.036eV above the ground state,
corresponding to the difference in ionization potentials of Rb and Yb.
Considering the long-range numerical values of the potential, we find that the hyperpolarizabilities
  in the multipole expansion \eqref{mult} need to be taken into account.
   Little spectroscopic information about quadrupole or octupole moments for 
   the heavier alkali atoms exists as it is very difficult to precisely measure the oscillator strengths.
  For the higher-order multipole terms we obtained solutions first by fitting the data to equation \eqref{mult}.
  In this case, we chose to fix the dipole and quadrupole polarizabilities by their 
  experimental values $\alpha_d=319.2$a.u~\cite{molo74} and $C_6=6480$a.u.~\cite{mitr03} 
  respectively. We then fitted to our data points to obtain $C_8$ using a 
  least-squares fitting procedure. The results in the MRCI and FCI approximations 
  are given in Table~\ref{tab4} and~\ref{tab5} respectively. In both cases (MRCI and FCI) 
  the scattering length $a_s$ for the ground electronic state $X^1\Sigma^+$ is shown. 
  The dipole polarizability for the ground state of Yb  from the MRCI and FCI approximations are respectively, 
  128.5 and 128.4 (c.f. Table \ref{tab3}).

For two electrons outside a closed shell one expects both the MRCI and FCI 
results to be in suitable agreement with each other. Fig 1 illustrated only the MRCI results as 
on the scale shown the FCI results are identical.
From our {\it ab initio} results while it is seen that the equilibrium 
bond distance $r_e$ and the dissociation energy $D_e$ are practically 
identical in the MRCI and FCI approximations, the scattering lengths are widely different. 
This we attribute to the difficulty of accurately obtaining the dispersion coefficients 
from the long range tail of the potential (1) in both approximations, which are subsequently 
used in numerically solving the radial Schr\"odinger equation (3) for the phase shift and corresponding 
scattering length.  We illustrate this fact by the values tabulated in Table \ref{tab4} and \ref{tab5} for the
scattering length.

\begin{table*}
\caption{Molecular constants and multipole coefficients obtained from the {\it ab initio} data for the lowest 
               three states of  ${\rm YbRb}^+$ in the MRCI approximation. The equilibrium bond length 
		$r_e$ and the dissociation energy $D_e$ were obtained using 
		a spline fit to the data.  $C_8$ and $E_\infty$ were obtained using a least-squares fit. 
		The scattering length given by $a_s$ is in atomic units.}
\begin{ruledtabular}
\begin{tabular}{ccccccc}
Molecular & $r_e$ & $D_e$ &                       $C_8$                  &  $E_\infty$   &  $a_s$ \\
Symmetry & ($a_0$) & (eV) &   $10^8$ a.u.  & (eV) &($a_0$) \\
\hline
X$^1\Sigma^+$ & 9.024 & 0.2244 &  	-		&  0 			& 2815\\
A$^1\Sigma^+$ & 14.179 & 0.1161 &  1.770 & 2.036 & 6766 \\
a$^3\Sigma^+$ & 10.135 & 0.6785 &  1.786 & 2.036 & 9646 \\
\end{tabular}
\end{ruledtabular}

\label{tab4}
\end{table*}

\begin{table*}
\caption{Molecular constants and multipole coefficients obtained from the {\it ab initio} data for the lowest 
                three states of  ${\rm YbRb}^+$  in the FCI approximation. The equilibrium bond length 
		$r_e$ and the dissociation energy $D_e$ were obtained using 
		a spline fit to the data.  $C_8$ and $E_\infty$ were obtained using a least-squares fit.
		The scattering  length given by $a _s$ is in atomic units.}
\begin{ruledtabular}
\begin{tabular}{ccccccc}
Molecular & $r_e$ & $D_e$ &                       $C_8$                  &  $E_\infty$   &  $a_s$ \\
Symmetry  & ($a_0$) & (eV) &   $10^8$ a.u.  & (eV) &($a_0$) \\
\hline
X$^1\Sigma^+$  & 9.031 & 0.2227 & 	-		&  0 			& $-11594$ \\
A$^1\Sigma^+$ & 14.184 & 0.1150 &   0.7020 & 2.035 & $-59036$ \\
a$^3\Sigma^+$  & 10.134 & 0.6776 &   1.2172 & 2.035 & $-3606$ \\
\end{tabular}
\end{ruledtabular}

\label{tab5}
\end{table*}

\section{Conclusions}
 
We have investigated the electronic structure of low lying states of the diatomic molecular ionic system 
containing a ytterbium ion and a rubidium atom, with relevance to ultracold chemistry , in particular,  charge transfer 
processes involving Yb ions and Rb atoms. 
We employed both a multi-reference configuration interaction (MRCI) and a full configuration interaction (FCI) 
approach to obtain turning points, crossing points, potential minima and spectroscopic molecular constants 
 for the  lowest five molecular states.  Long-range parameters, including  
the dispersion coefficients are estimated from our {\it ab initio} data.
We find a near degeneracy of the Yb$^{+}$ ground state with excited charge transfer channels.
The results for the long-range potential tail including the polarizability 
are seen to be in suitable agreement with previous experimental and theoretical work.
We present preliminary results for the ultracold elastic scattering amplitude for the
ytterbium ion colliding with the rubidium atom, based on our {\it ab initio}  data including both asymptotic ionic channels.  
These estimates, assuming pure adiabatic elastic scattering, 
indicate that the Yb$^{+}$ ion collisions with Rb atom interactions are attractive for both the singlet and triplet states.
Effective range theory  is used to derive the corresponding pseudopotential which has a strong energy 
dependence, even  in the nK r\'egime not well represented by the single-parameter scattering length. 
 The well-known  sensitivity of scattering length to changes in the potential indicates that our estimates, although 
 (to our knowledge) the first of their kind,  are rather crude. 
 Nonetheless, the potential energy curves are accurate in the region where charge exchange is important and 
 thus offer prospects of studying this process under quantum controlled conditions.

 \vspace*{-.5cm}
\begin{acknowledgments}
\vspace*{-.25cm}
HDLL is grateful to the Department of Employment and Learning (Northern Ireland) for the provision of a
postgraduate studentship. J.G. would like to acknowledge funding from an IRCSET Marie Curie International Mobility fellowship.
B MMcL would like to thank Queen's University Belfast for the award of a Visiting Research Fellowship during 
which this work was performed. Part of the computational work was carried out at the 
National Energy Research Scientific Computing Center in Oakland, CA, USA.

\end{acknowledgments}
 

\end{document}